\documentclass[twocolumn,amsmath,amssymb]{revtex4}

\usepackage[pdftex]{graphicx}

\usepackage[breaklinks]{hyperref}
\hypersetup{
    bookmarks=false,
    pdfstartview={FitH},
    colorlinks=true,
    linkcolor=blue,
    citecolor=blue,
    urlcolor=blue
}

\begin{document}

\title{Maximum entropy deconvolution of resonant inelastic x-ray scattering
spectra}

\author{J.~Laverock}
\author{A.~R.~H.~Preston}
\author{D.~Newby,~Jr.}
\author{K.~E.~Smith}
\affiliation{Department of Physics, Boston University, 590 Commonwealth Avenue,
Boston, Massachussetts, MA 02215, USA}
\author{S.~B.~Dugdale}
\affiliation{H.~H.~Wills Physics Laboratory, University of Bristol, Tyndall
Avenue, Bristol BS8 1TL, United Kingdom}

\begin{abstract}
Resonant inelastic x-ray scattering (RIXS) has become a powerful tool in the
study of the electronic structure of condensed matter. Although the linewidths
of many RIXS features are narrow, the experimental broadening can often
hamper the identification of spectral features.
Here, we show that the Maximum
Entropy technique can successfully be applied in the deconvolution of RIXS
spectra, improving the interpretation of the loss features without a severe
increase in the noise ratio.
\end{abstract}

\maketitle

\section{Introduction}
Over the last decade or so, the technique of resonant inelastic soft x-ray
scattering (RIXS) has developed into a formidable tool in the study of the
electronic structure of the solid state \cite{kotani2001,ament2011b},
nanomaterials
\cite{liu2007}, and even liquids and gases \cite{guo2003,hennies2010}. The
strengths of RIXS using soft x-rays owe to its bulk sensitivity, atomic
(and even orbital) selectivity, ability to sample finite momentum transfer
and narrow linewidth.  Indeed, the recent development of
sub-100~meV resolution grating spectrometers \cite{ghiringhelli2009b}
has opened up the new possibility of using soft x-rays to study low-energy
collective excitations such as magnons or orbitons, and even to track their
dispersion in momentum \cite{schlappa2009,braicovich2009}.

In the RIXS process, a core electron is excited into an unoccupied valence
orbital, creating a core hole in the intermediate state which then rapidly (and
coherently) decays to the final state via x-ray emission (see, for example,
Refs.~\cite{kotani2001,ament2011b}). The incident x-ray is tuned near the
threshold of a particular core electron excitation, granting RIXS its site
and orbital selectivity. For example, for transition metal $L$-edges the
process is of the form, $2p^6\,3d^n \rightarrow 2p^5\,3d^{n+1} \rightarrow
2p^6\,3d^{n*}$, in which the * denotes an excited state. The final state
may be an excited configuration (e.g.\ $dd^*$ crystal-field transition)
or a collective excitation, and the energy difference between the incident
and emitted x-rays represents the excitation energy.

Typically however, for soft x-ray RIXS where charge-transfer and/or
crystal-field excitations are of interest, the combined energy resolution of
the incident photons and RIXS spectrometer amounts to 0.5~--~1.0~eV
near the O $K$-edge ($\sim 520$~eV). On the other hand, the typical energy of
transition metal $dd^*$ crystal field excitations is of the order of 1~--~4~eV,
and their separation can be close to the limit of the resolving power of
moderate-resolution instruments (for example, low-energy $dd^*$ transitions in
VO$_2$ \cite{piper2010b,braicovich2007}).  Moreover, even in high-resolution
measurements, different low-energy collective excitations can lie in close
proximity relative to the resolution function; for example, bi- and single
magnon peaks as well as phonon contributions within 400~meV in La$_2$CuO$_4$
\cite{braicovich2010}.  There is therefore a sensitive trade-off between
statistical precision and resolution for these kinds of measurements,
which are perfectly poised to take advantage of the benefits of a reliable
deconvolution procedure.

The technique of Maximum Entropy (MaxEnt) owes its origins to the study
of communication theory introduced by Shannon \cite{shannon1949}, in which
the proposed measure of information content, $S$, had the same form as the
thermodynamic entropy,
\begin{equation}
S = - k \sum_{i,j} p_{i,j} \ln p_{i,j},
\label{e:entropy}
\end{equation}
in which $p_{i,j}$ is the number of counts in a pixel $(i,j)$ and $k$ is
an arbitrary constant. The basic idea of MaxEnt is relatively simple:
one maximizes the information content, $S$, of the processed (deconvoluted)
probability distribution subject to it being consistent with the measured data.
This consistency test is achieved through a $\chi^2$ comparison between the
processed distribution, convoluted with the (known) resolution function,
and the measured distribution. The outcome of this process is a distribution
that is `most likely' to have been responsible for the measured data, given
the properties of the resolution function.

The algorithm employed here is the Cambridge Algorithm \cite{gull1985} and
has been successfully applied to many fields of data analysis, e.g.~positron
annihilation \cite{dugdale1994,fretwell1995}, image analysis, astrophysics and
extended x-ray absorption fine structure data \cite{gull1984}.
The solution of the algorithm is iteratively updated in three `search
directions' $d_{i,j,n}$, with coefficients $\alpha_n$:
\begin{equation}
p^{\prime}_{i,j} = p_{i,j} + \sum_{n=1}^3 \alpha_n d_{i,j,n}.
\end{equation}
The search directions used in this algorithm are $\nabla{S}$ (to maximize the
entropy), $\nabla{\chi^2}$ (to minimize the $\chi^2$), and a third direction
involving higher derivatives. The solution, $p^{\prime}_{i,j}$, and
coefficients $\alpha_n$ are updated at each iteration until convergence in the 
solution is achieved.
The particular benefit
of the MaxEnt procedure over more traditional Fourier transform-based
deconvolution methods is the substantially improved signal-to-noise ratio (which
will be discussed in more detail in Section \ref{s:noise}), even
when faced with sparse (or missing) data \cite{gull1984}.

In the following, all measurements were performed at the AXIS endstation
of beamline 7.0.1 of the Advanced Light Source, Lawrence Berkeley National
Laboratory, employing a Nordgren-type grating spectrometer \cite{nordgren1989}.
Spectra were
recorded on a 2D multi-channel plate detector.
Correction for the curvature introduced by the optical components of
the spectrometer was achieved by fitting the peak position of several
well-characterized sharp emission features across the non-dispersive axis
of the detector. This correction was then used for all subsequent spectra.
The MaxEnt procedure
was applied to the raw 2D spectra before curvature correction. This was
found to produce more favorable results compared with processing corrected,
integrated 1D spectra, a finding that is expected owing to the greater
information content of the 2D images. The MaxEnt deconvolution process
involves convolving the processed data with the known resolution function
(the `broadening' function), for which there are two components in RIXS
measurements: i)~the energy resolution of the incident photons,
and ii)~the spectrometer resolution. The second of
these, predominantly arising from the finite source size, is much more easily
dealt with: it is approximately Gaussian with respect to the wavelength of
the emitted photons. An accurate knowledge of the incident photon resolution is
much more challenging since its impact on emission spectra is non-trivial
and depends in part on the specific excitations involved in the vicinity of
the incident photon energy (i.e.~the absorption spectrum); no attempt has
been made to remove this component from the experimental data.

\begin{figure}[t!]
\begin{center}
\includegraphics[width=1.00\linewidth,clip]{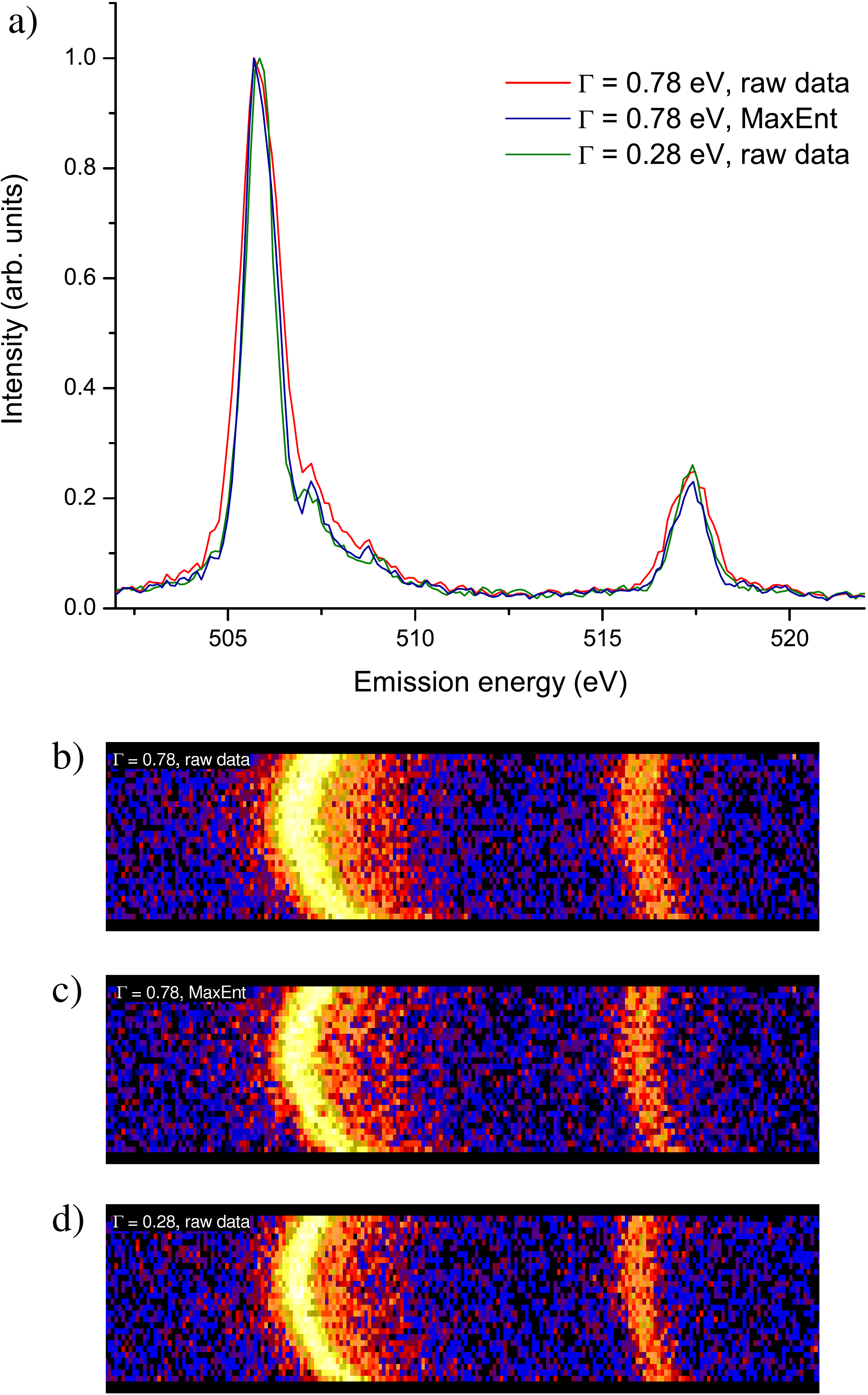}
\end{center}
\caption{\label{f:zn_ref} (Color online) Comparison between raw (second order)
Zn x-ray emission spectra recorded at 0.78~eV spectrometer resolution and the
results of the MaxEnt deconvolution procedure. For comparison, a spectrum
recorded at 0.28~eV spectrometer resolution is also shown. a) Integrated
spectra, b) raw 2D spectrum for $\Gamma = 0.78$~eV, c) MaxEnt deconvoluted
2D spectrum for $\Gamma = 0.78$~eV, and d) raw 2D spectrum for $\Gamma =
0.28$~eV. The raw 2D spectra (before correcting for the curvature of the image)
are shown on a logarithmic false color scale.}
\end{figure}

\section{Deconvolution of reference spectra}
As an initial diagnosis of the performance of the MaxEnt procedure, x-ray
emission spectra of the $L_{3,2}$-edge of Zn were obtained at various different
spectrometer slit widths, corresponding to different spectrometer resolutions.
These spectra represent transitions of the form $3d \rightarrow 2p$ with
the spin-orbit split $L_3$ edge at a lower energy than the $L_2$ emission
line.  Since we are dealing with emission features, excited well above the
absorption threshold, the incident photon energy does not contribute to
the overall resolution of these measurements, and they therefore provide
a robust test of the MaxEnt procedure in removing the spectrometer part
of the resolution function.  In Fig.~\ref{f:zn_ref}, Zn spectra recorded
(in second order \cite{secondorder})
with spectrometer resolutions of 0.28~eV and 0.78~eV at
FWHM are shown, alongside the results of the MaxEnt deconvolution of the
0.78~eV spectrum. A broadening function of 0.66~eV (85\% of the spectrometer
resolution) was used in the MaxEnt deconvolution routine. It is clear
from Fig.~\ref{f:zn_ref} that the width of the deconvoluted spectrum is
much narrower than the raw spectrum, and indeed approaches the width of the
narrower 0.28~eV spectrum, indicating that a large portion of the instrument
resolution has been removed from the spectrum. In fact, this narrowing of the
line width of the emission lines is already visible in the raw 2D spectra
shown in Fig.~\ref{f:zn_ref}b-d. The FWHM of the $L_3$ edge (including the
natural width of this feature) is 1.29~eV for the raw spectrum, compared with
1.06~eV after deconvolution. For comparison, the FWHM of the same feature
in the narrower raw spectrum is 1.00~eV. Moreover, there are no additional
artefacts introduced in the deconvolution procedure -- the MaxEnt spectrum
closely resembles the 0.28~eV spectrum. The behavior of the deconvolution
with varying broadening functions was also investigated, and found to
be very stable for functions of FWHM $\leq 90$\% of the total resolution
function. Above this, some artificial sharpening close to the emission peaks
was observed. Finally, it is worth noting that the signal-to-noise ratio is
not significantly decreased in the deconvoluted spectra (a subject to which we
will return in Section \ref{s:noise}).

\section{Deconvolution of crystal field excitations}
In order to test its performance in resolving close spectral features,
the MaxEnt deconvolution procedure was applied to Co $L$-edge RIXS data
of Co$_3$V$_2$O$_8$, which is a kagom\'{e} staircase compound consisting
of Co$^{2+}$O$_6$ octahedra separated by V$^{5+}$O$_4$ tetrahedra
\cite{balakrishnan2004}. Spectra were recorded in second order with a
spectrometer resolution of 0.82~eV and an incident photon
resolution of 0.4~eV, amounting to a total resolution of approximately
0.91~eV. For the MaxEnt deconvolution procedure, a broadening function
of 0.82~eV was used, equivalent to the spectrometer resolution.  The same
procedure was applied to Co $L$-edge emission reference spectra in order to
check that no artefacts were introduced in the deconvolution.

\begin{figure}[t!]
\begin{center}
\includegraphics[width=1.00\linewidth,clip]{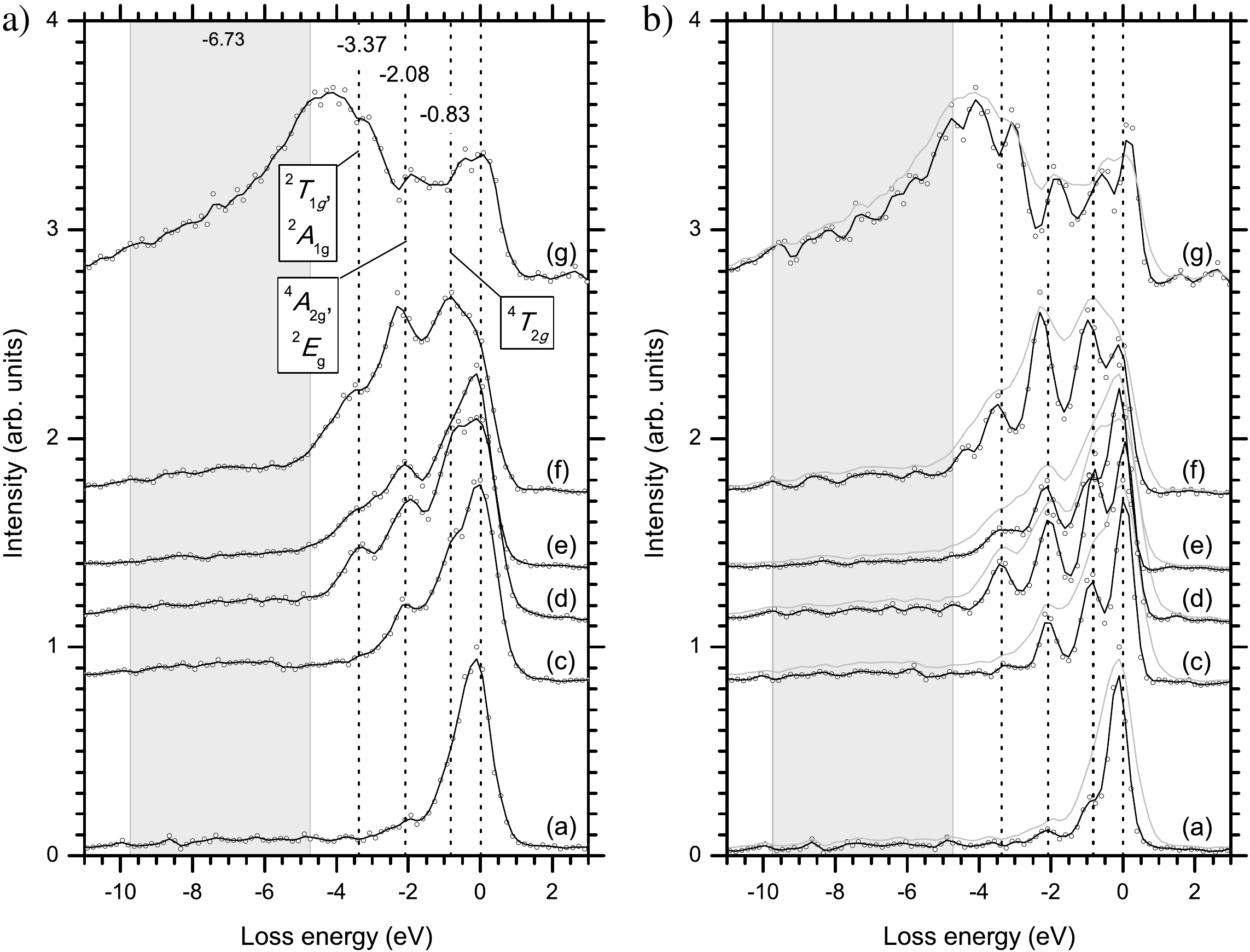}
\end{center}
\caption{\label{f:cvol} Co $L$-edge RIXS data of Co$_3$V$_2$O$_8$. (a) Raw
spectra, and (b) MaxEnt deconvoluted spectra, in which the raw spectra are
reproduced in light grey. Solid lines are a guide to the eye and represent the
data with a binomial smoothing of 0.15~eV. Vertical lines show the
elastic line at 0~eV and loss features identified in the raw data. The symmetry
of the excitations are labelled in (a).}
\end{figure}

The raw RIXS data are shown on a loss energy scale in Fig.~\ref{f:cvol}a
for several different incident photon energies between 777~eV (spectrum a)
and 782~eV (spectrum g), spanning the Co $L_3$-edge absorption feature. The
peak at 0~eV represents elastically scattered light. There are clear loss
features present in all spectra below $-4$~eV, above which a weak and broad
charge-transfer peak emerges centered about $-6.7$~eV.  In order to accurately
locate these peaks, a linear combination of Voigt functions was fitted
to each spectrum, and the average center of each feature was determined.
Altogether, three distinct loss features can be seen in the spectra: (i) at
$-0.83$~eV clearly visible in spectra (d) and (f), (ii) at $-2.08$~eV in spectra
(c-f), and at $-3.37$~eV in spectrum (d). These spectra, and the energies of
these features, are similar to RIXS $L$-edge measurements of other Co$^{2+}$
compounds, for example CoO \cite{chiuzbaian2008}.

The MaxEnt deconvoluted spectra are shown in Fig.~\ref{f:cvol}b, and show
obvious improvement in the linewidth of the features. In these spectra, the loss
features previously identified are now much more clearly visible. For example,
features (i) and (ii) are directly visible in all spectra, and (iii) is visible
in all but the lowest energy spectrum (a). It is emphasized that no attempt to
re-determine the energy location of these features with the new information
provided by the MaxEnt algorithm has been made, yet they agree within $\sim
0.1$~eV in all spectra. It is noted that the noise level of the MaxEnt
spectra is slightly higher than for the raw data, as might be expected in any
deconvolution procedure. Nevertheless, focussing on features that persist across
several incident photon energies allows one to be sure that the origin of a
feature is intrinsic to the system under study, and not an artefact of the
increased noise floor. The strength of this approach is anticipated to
lie when spectral features are difficult to identify in raw data, as for example
in spectrum (c), in which the $-0.83$~eV feature is hard to separate from the
elastic peak.

\begin{figure}[t!]
\begin{center}
\includegraphics[width=1.00\linewidth,clip]{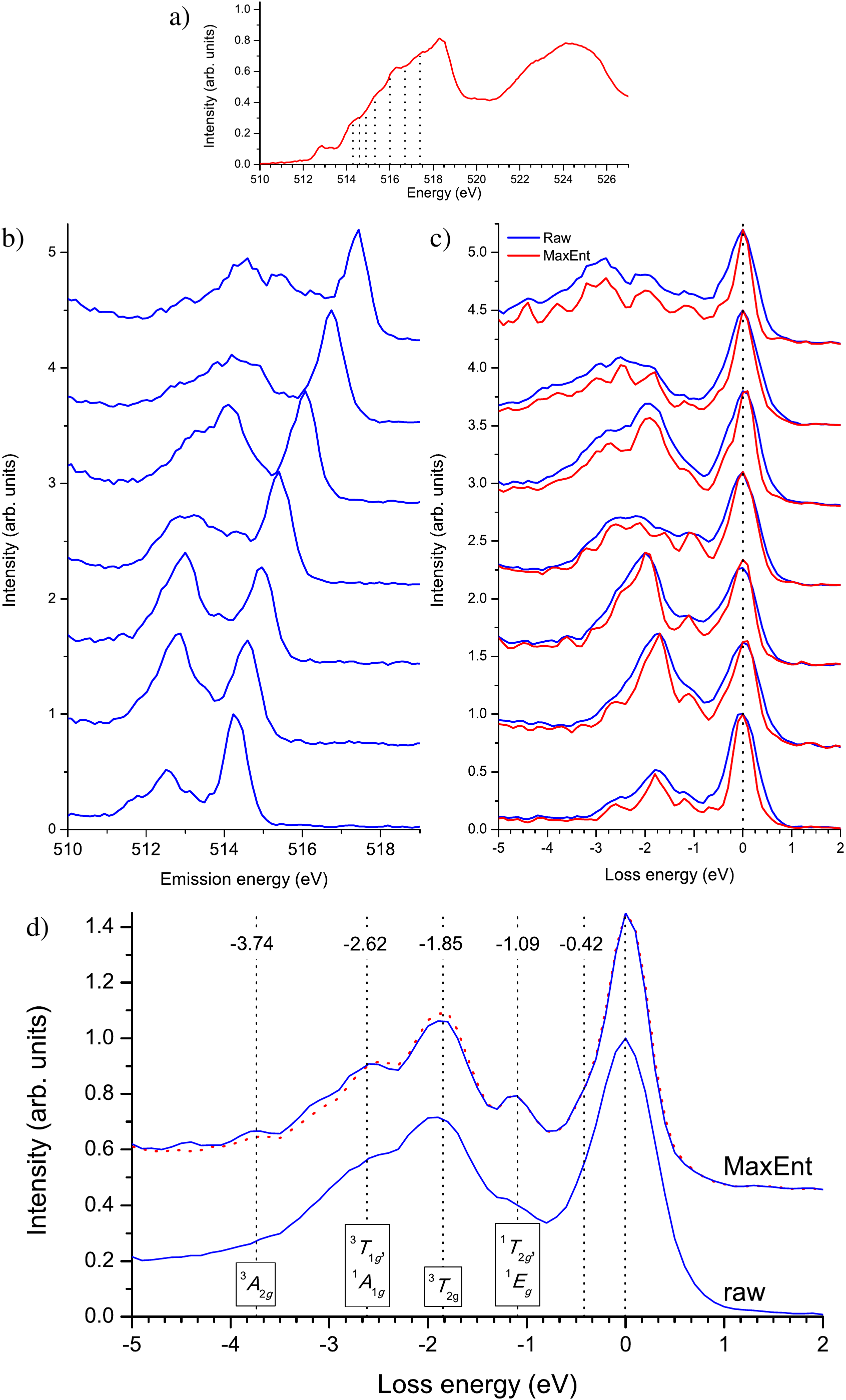}
\end{center}
\caption{\label{f:rvo3} (Color online) V $L$-edge RIXS data of NdVO$_3$. (a)
The absorption spectrum and excitation energies used for RIXS. Raw RIXS data
are shown on (b) an emission energy, and (c) a loss energy scale, for which
the individual MaxEnt spectra are also shown. (d) Summed RIXS spectra before
and after MaxEnt deconvolution. The symmetry of the excitations are labelled
in (d); the dotted line represents the MaxEnt summed data with one spectrum
missing (see text).}
\end{figure}

\section{Deconvolution of elastic peaks}
As an additional test, the MaxEnt procedure has been applied to V $L_3$-edge
RIXS data of NdVO$_3$ at room temperature, a system in which the occupation of
the V $3d$ orbitals becomes ordered at
low temperature \cite{miyasaka2003}. Spectra were recorded with a spectrometer
and incident photon resolution of 0.36~eV, yielding a total energy resolution of
around 0.51~eV. Seven spectra were obtained approximately equally-spaced
across the V $L_3$-edge absorption feature (and are shown in Fig.~\ref{f:rvo3}),
and each spectrum was treated to
the MaxEnt procedure with a broadening function of 0.36~eV. Again, the same
procedure was applied to a Zn $L$-edge emission reference spectrum to ensure
no artefacts were introduced in the deconvolution. The relative
intensity of RIXS spectra at the V $L$-edge is weak, and so the spectra
here have been summed together to yield `averaged' loss spectra, shown in
Fig.~\ref{f:rvo3}d for both the raw and MaxEnt deconvoluted spectra. The idea
behind this procedure is that fluorescent features, that are dispersive in loss
energy, contribute a broad and weak background, whereas loss features will
reinforce in the summed spectra. Note that there will be a slight additional
broadening of the features due to uncertainty in the initial photon energy.

The individual raw spectra are shown in Fig.~\ref{f:rvo3}b.  The first spectrum
is on the onset of the V $L_3$-edge, in a location most suitable for exploring
loss features. At higher excitation energies, V $3d$ fluorescence begins to
contribute more strongly in the spectra, but is mostly concentrated beyond
2.5~eV from the elastic peak. Therefore, the subsequent interpretation of
features in the summed spectra below this energy is less influenced by the
presence of fluorescence. Moreover, the contributions from fluorescence do
not reinforce across spectra, and provide a weak background to the summed
spectra presented in the manuscript.
In the raw summed spectra (Fig.~\ref{f:rvo3}d), two principle features
are present: the elastic peak at O~eV and a broader feature at around
$-2$~eV. Shoulders either side of this second peak indicate the presence of
other spectral features. Once the MaxEnt procedure is applied, however,
the location of these weaker features becomes clear, at $-2.62$~eV and
$-1.09$~eV either side of the $-1.85$~eV peak, and represent crystal field
$dd^*$ excitations.  These features are also clearly identifiable in the
individual deconvoluted spectra shown in Fig.~\ref{f:rvo3}c, particularly
the lowest energy spectrum.
Furthermore, however, a relatively weak shoulder is
evident close to the elastic peak at $-0.42$~eV, and may represent an orbital
excitation in the form of a bi-orbiton, previously observed for YVO$_3$
at 0.4~eV \cite{benckiser2008}.  Its presence can also be inferred in the
raw spectra from the slight asymmetry of the elastic peak in this spectrum.
Finally, an additional feature at $-3.74$~eV, not visible in the raw spectrum,
becomes clear after the MaxEnt deconvolution. This peak is harder to identify in
the individual spectra, but is most prominant in the highest energy MaxEnt
spectrum in Fig.~\ref{f:rvo3}c. In order to check its presence at more than one
excitation energy (and ensure that its origin is not the fluorescent part of the
spectrum), the highest energy MaxEnt spectrum has been removed from the
summation in the dotted line of Fig.~\ref{f:rvo3}d, and indeed, this peak still
persists. However, a higher energy feature (at around $-4.5$~eV) disappears in
this process; although this may be a loss feature of the spectrum, further work
is needed to confirm its origin.

The strong elastic peak in these data mean that we can directly measure
the instrument resolution, since the elastic peak is a $\delta$-function in
the limit of an infinitessimally small total resolution function. In order to
avoid complications with low-energy (e.g.\ phonon) excitations, all the
following fits to the elastic peak have been constrained predominantly to its
high-energy side. For the
raw spectra, the FWHM of the elastic peak is 0.75~eV,
slightly broader than expected, presumably in part due to the additional
processing involved in summing these spectra (but possibly also due to some
uncertainty in the slit width). For the MaxEnt spectra, however, the FWHM is
0.50~eV. Although at first glance this is not as narrow as one might expect,
it should be remembered that this represents a combination of the incident
photon resolution and the spectrometer resolution, and no attempt has
been made to remove the incident photon resolution function.  In this respect,
the
narrowing of the elastic peak behaves well, and corresponds to a narrowing
of the `effective' spectrometer resolution (by subtracting in quadrature
the incident photon resolution, 0.36~eV, from the measured resolution) from
0.66~eV to 0.35~eV.

\section{Noise Propagation}
\label{s:noise}
\begin{figure}[h!]
\begin{center}
\includegraphics[width=0.95\linewidth,clip]{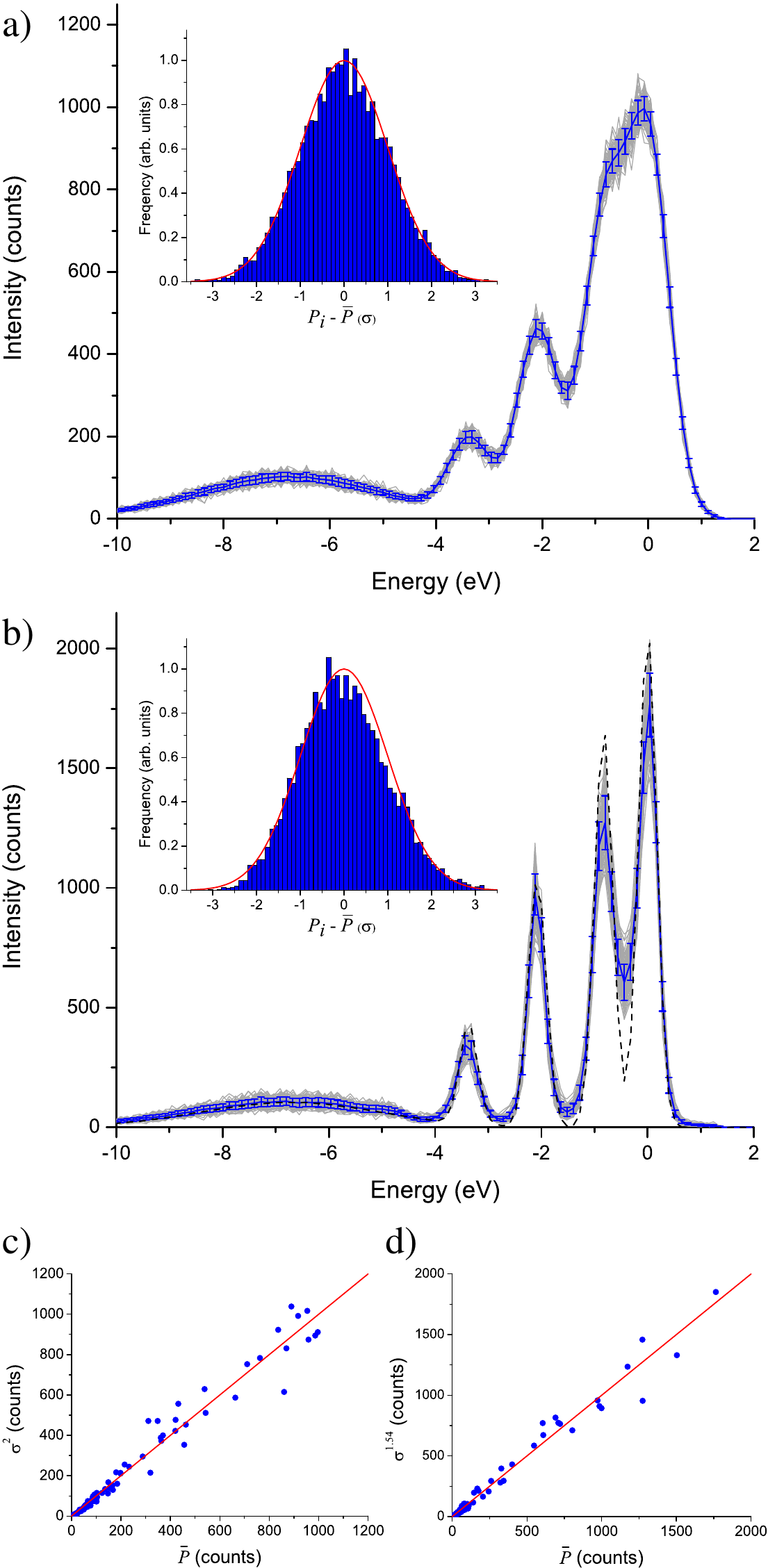}
\end{center}
\caption{\label{f:noise} (Color online) The results of deconvoluting $M=100$
simulated noisy spectra with the MaxEnt procedure; (a) Input spectra, and (b)
deconvoluted spectra. The individual spectra, $P_i(E)$, are shown in light
gray, with the mean spectra, $\bar{P}(E)$, in blue. The error bars shown
represent the standard deviation, $\sigma[P(E)]$, of the $M$ spectra for each
energy. The inset shows the distribution of each simulated spectrum from its
mean, normalized to its standard deviation. In (b), the initial test spectrum,
$T(E)$, is shown by the dashed line. (c) Comparison between the mean of the
input spectra and their variance, $\sigma^2$.  (d) Comparison between the mean
of the deconvoluted spectra and $\sigma^{1.54}$.  For (c) and (d) the solid line
represents $\bar{P} = \sigma^2$ and $\bar{P} = \sigma^{1.54}$ respectively.}
\end{figure}

The propagation of noise through the MaxEnt procedure is a complex problem;
indeed, with only a slightly different setup of the problem, MaxEnt can be
used to `de-noise' data or reconstruct missing information from sparse data
\cite{skilling1991,gull1984}. In order to quantify the propagation of noise
through the MaxEnt procedure, we have simulated a series (of $M = 100$) of
noisy RIXS spectra, $P_i(E)$.  These have then been passed through the MaxEnt
deconvolution, and the variance in the resulting deconvoluted spectra analyzed,
\begin{equation}
\sigma^2[P(E)] = \frac{1}{M} \sum_{i=1}^{M} [P_i(E) - \bar{P}(E)]^2,
\end{equation}
where $\bar{P}(E)$ is the mean of the $M$ simulated spectra. The test
spectrum, $T(E)$, was chosen to approximate spectrum (d) in Fig.~\ref{f:cvol}:
a linear combination of Gaussian functions of 0.4~eV FWHM (similar to the
beamline resolution used in the measurements) centered at 0~eV, $-0.83$~eV,
$-2.08$~eV and $-3.37$~eV were added to a Gaussian function of 4~eV FWHM
centered at $-6.73$~eV to approximate the elastic, $dd^*$ and CT features
respectively. This spectrum was then convoluted with a Gaussian function of
0.82~eV FWHM, approximating the spectrometer resolution, and scaled to contain
1000~counts in the peak data channel. The $M$ simulated spectra were then
created by adding randomly generated noise following a normal distribution with
$\sigma(E) = \sqrt{T(E)}$, and are shown in Fig.~\ref{f:noise}a.  These test
spectra, $P_i(E)$, were each deconvoluted with the same parameters as used
for the Co$_3$V$_2$O$_8$ data shown in Fig.~\ref{f:cvol}.

The resulting deconvoluted spectra are shown in Fig.~\ref{f:noise}b.
Comparison between the deconvoluted spectra and the initial test spectrum,
$T(E)$ [shown by the dashed line in Fig.~\ref{f:noise}b], are very encouraging,
and provide a direct visualization of the power of the MaxEnt algorithm.
In order to ensure that the mean and variance are meaningful quantities,
the inset shows the distribution of the data about the mean, in units of
the standard deviation, for data between $-9$ and 1~eV (to restrict the
contribution to the finite signal region of the spectra).  Although there
is a weak positive skewness in the distribution of the deconvoluted spectra
(emphasizing the complexity of the MaxEnt noise problem), it is sufficiently
close to a normal distribution that the mean and variance are still useful
indicators of the distribution. This is important since it allows us to
attach meaningful (statistical) errors to the deconvoluted spectra.

\begin{figure}[t!]
\begin{center}
\includegraphics[width=1.00\linewidth,clip]{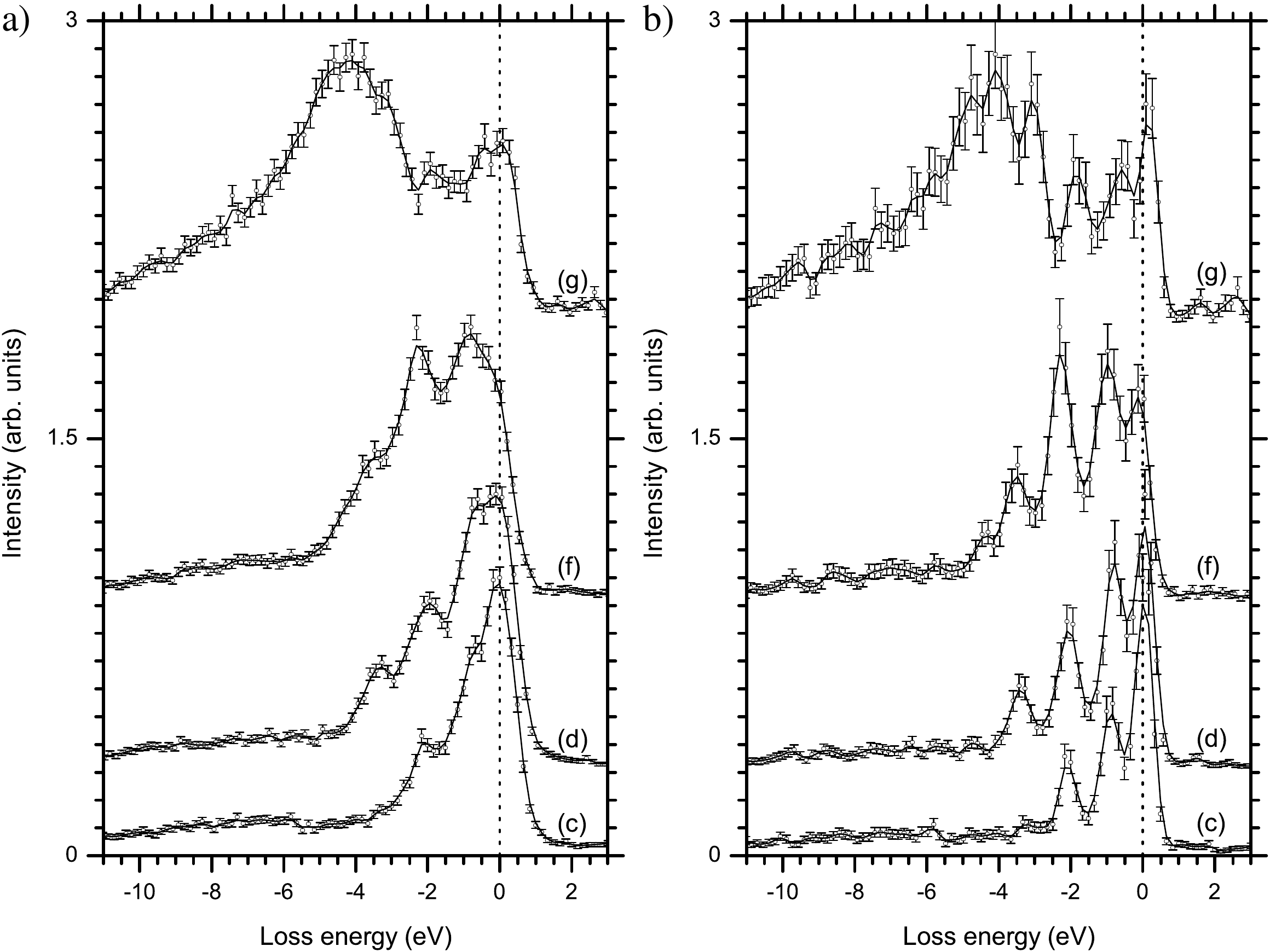}
\end{center}
\caption{\label{f:cvolerr} (Color online) The (a) raw, and (b) MaxEnt
deconvoluted RIXS spectra shown in Fig.~\ref{f:cvol}, reproduced here with
error bars. In (a), the statistical error is shown ($\sigma = \sqrt{N}$),
whereas in (b) the empirical error $\sigma = N^{1/1.54}$ is used (see text).}
\end{figure}

It is evident from Figs.~\ref{f:noise}a,b that, as expected, the noise
level is slightly higher for the deconvoluted spectra (for example, compare
the error bars for the peak at 0~eV in Fig.~\ref{f:noise}a and at $-2$~eV
in Fig.~\ref{f:noise}b, for which the intensity is similar). For Poisson
statistics, relevant in counting problems, the variance of a datum scales with
its expected value (for $\bar{x} \gtrsim 10$), as shown in Fig.~\ref{f:noise}c
in which the variance, $\sigma^2$, of the input data is plotted against its
mean, $\bar{P}$. However, this is not the case for the deconvoluted spectra,
indicating the propagation of errors through the MaxEnt process is non-linear
with respect to the pixel intensity. Qualitatively, the relationship is super
linear, meaning that pixels of high intensity (in the deconvoluted spectra)
are more sensitive to statistical noise than those of lower intensity. Such
behavior is connected with the tendency of the deconvolution to move counts
from low intensity regions of the input spectrum towards higher intensity
regions. Empirically, we find that the variance and mean are connected by
$\bar{P} \sim \sigma^{1.54}$ for the range of $\bar{P}$ investigated here
(approximately 50~--~2000), as shown in Fig.~\ref{f:noise}d, which corresponds
to an approximate doubling of the noise ratio for a typical spectrum.  It is
emphasized that this analysis only reflects the propagation of statistical
noise, and does not account for systematic errors that may be present
in the process itself.  The result of applying these empirical errors to
the Co$_3$V$_2$O$_8$ RIXS data is shown in Fig.~\ref{f:cvolerr} for some
representative spectra before and after the MaxEnt procedure. In each case,
the visual scatter of the data points is consistent with the magnitude of the
error bars, and the loss features that were previously identified are well
above the noise level. Moreover, the apparent structure at high energies in
the deconvoluted spectrum (g) is of the order of the noise, and is due to
the poorer statistics recorded for this spectrum.

\section{Conclusions}
In summary, the MaxEnt deconvolution procedure has been successfully
applied to soft x-ray RIXS spectra. The deconvoluted spectra show a marked
improvement in the resolution of spectral features without introduction of
artefacts associated with the process or excessive increase in the noise
ratio, allowing for greater confidence in separating and identifying loss
features, such as crystal-field $dd^*$ transitions or low-energy collective
excitations. For example, a very recent application of MaxEnt has helped to
clarify the RIXS features of La$_{1-x}$Sr$_x$MnO$_3$ \cite{piper2011b}.
Detailed analysis of the propagation of noise through the MaxEnt procedure has
been presented, and the noise ratio of deconvoluted spectra has been found to
approximately double for the features of typical spectra.
The process is quite general, and is not limited to the soft
x-ray regime or 2D data, and is expected to perform equally well with, for
example, x-ray absorption spectra or high-resolution RIXS. 
It is anticipated that the strength of
the procedure is when spectral features are difficult to identify in raw data
(but whose presence may already be inferred from that data, albeit indirectly).

\section*{Acknowledgements}
The Boston University (BU) program is supported in part by the Department of
Energy under Contract No. DE-FG02-98ER45680.  The Advanced Light Source is
supported by the Director, Office of Science, Office of Basic Energy Sciences,
of the U.S. Department of Energy under Contract No. DE-AC02-05CH11231.


\begin{thebibliography}{99}

\bibitem{kotani2001}
A.\ Kotani and S.\ Shin,
\href{http://dx.doi.org/10.1103/RevModPhys.73.203}
{Rev.\ Mod.\ Phys.\ {\bf 73}, 203 (2001)}.

\bibitem{ament2011b}
L.\ J.\ P.\ Ament, M.\ van Veenendaal, T.\ P.\ Devereaux, J.\ P.\
Hill and J.\ van den Brink,
\href{http://dx.doi.org/10.1103/RevModPhys.83.705}
{Rev.\ Mod.\ Phys.\ {\bf 83}, 705 (2011)}.

\bibitem{liu2007}
H.\ Liu, J.-H.\ Guo, Y.\ Yin, A.\ Augustsson, C.\ Dong, J.\ Nordgren, C.\ Chang,
P.\ Alivisatos, G.\ Thornton, D.\ F.\ Ogletree, F.\ G.\ Requejo, F.\ de Groot
and M.\ Salmeron,
\href{http://dx.doi.org/10.1021/nl070586o}
{Nano Lett.\ {\bf 7}, 1919 (2007)}.

\bibitem{guo2003}
J.-H.\ Guo, Y.\ Luo, A.\ Augustsson, S.\ Kashtanov, J.-E.\ Rubensson, D.\ K.\
Shuh, H.\ {\AA}gren and J.\ Nordgren,
\href{http://dx.doi.org/10.1103/PhysRevLett.91.157401}
{Phys.\ Rev.\ Lett.\ {\bf 91}, 157401 (2003)}.

\bibitem{hennies2010}
F.\ Hennies, A.\ Pietzsch, M.\ Berglund, A.\ F\"{o}hlisch, T.\ Schmitt, V.\
Strocov, H.\ O.\ Karlsson, J.\ Andersson and J.-E.\ Rubensson,
\href{http://dx.doi.org/10.1103/PhysRevLett.104.193002}
{Phys.\ Rev.\ Lett. {\bf 104}, 193002 (2010)}.

\bibitem{ghiringhelli2009b}
G.\ Ghiringhelli, A.\ Piazzalunga, X.\ Wang, A.\ Bendounan, H.\ Berger, F.\
Bottegoni, N.\ Christensen, C.\ Dallera, M.\ Grioni, J.-C.\ Grivel, M.\ Moretti
Sala, L.\ Patthey, J.\ Schlappa, T.\ Schmitt, V.\ Strocov and L.\ Braicovich,
\href{http://dx.doi.org/10.1140/epjst/e2009-00993-8}
{Eur.\ Phys.\ J.\ Special Topics {\bf 169}, 199 (2009)}.

\bibitem{schlappa2009}
J.\ Schlappa, T.\ Schmitt, F.\ Vernay, V.\ N.\ Strocov, V.\ Ilakovac, B.\
Thielemann, H.\ M.\ R{\o}nnow, S.\ Vanishri, A.\ Piazzalunga, X.\ Wang, L.\
Braicovich, G.\ Ghiringhelli, C.\ Marin, J.\ Mesot, B.\ Delley and L.\ Patthey,
\href{http://dx.doi.org/10.1103/PhysRevLett.103.047401}
{Phys.\ Rev.\ Lett. {\bf 103}, 047401 (2009)}.

\bibitem{braicovich2009}
L.\ Braicovich, L.\ J.\ P.\ Ament, V.\ Bisogni, F.\ Forte, C.\ Aruta, G.\
Balestrino, N.\ B.\ Brookes, G.\ M.\ De Luca, P.\ G.\ Medaglia, F.\ Miletto
Granozio, M.\ Radovic, M.\ Salluzzo, J.\ van den Brink, and G.\
Ghiringhelli,
\href{http://dx.doi.org/10.1103/PhysRevLett.102.167401}
{Phys.\ Rev.\ Lett.\ {\bf 102}, 167401 (2009)}.

\bibitem{piper2010b}
L.\ F.\ J.\ Piper, A.\ DeMasi, S.\ W.\ Cho, A.\ R.\ H.\ Preston, J.\ Laverock,
K.\ E.\ Smith, K.\ G.\ West, J.\ W.\ Lu and S.\ A.\ Wolf,
\href{http://dx.doi.org/10.1103/PhysRevB.82.235103}
{Phys.\ Rev.\ B {\bf 82}, 235103 (2010)}.

\bibitem{braicovich2007}
L.\ Braicovich, G.\ Ghiringhelli, L.\ H.\ Tjeng, V.\ Bisogni, C.\ Dallera, A.\
Piazzalunga, W.\ Reichelt and N.\ B.\ Brookes,
\href{http://dx.doi.org/10.1103/PhysRevB.76.125105}
{Phys.\ Rev.\ B {\bf 76}, 125105 (2007)}.

\bibitem{braicovich2010}
L.\ Braicovich, J.\ van den Brink, V.\ Bisogni, M.\ Moretti Sala, L.\ J.\ P.\
Ament, N.\ B.\ Brookes, G.\ M.\ De Luca, M.\ Salluzzo, T.\ Schmitt, V.\ N.\
Strocov and G.\ Ghiringhelli,
\href{http://dx.doi.org/10.1103/PhysRevLett.104.077002}
{Phys.\ Rev.\ Lett.\ {\bf 104}, 077002 (2010)}.

\bibitem{shannon1949}
C.\ E.\ Shannon and W.\ Weaver, {\em The Mathematical Theory of Communication},
University of Illinois Press, (1949).

\bibitem{gull1985}
S.\ F.\ Gull and J.\ Skilling, {\em Maximum Entropy and Bayesian Methods in
Inverse Problems}, edited by C.\ R.\ Smith and W.\ T.\ Grandy, Dordrecht: Reidel
(1985).

\bibitem{dugdale1994}
S.\ B.\ Dugdale, M.\ A.\ Alam, H.\ M.\ Fretwell, M.\ Biasini and D.\ Wilson,
\href{http://dx.doi.org/10.1088/0953-8984/6/31/003}
{J.\ Phys.: Condens.\ Matter {\bf 6}, L435 (1994)}.

\bibitem{fretwell1995}
H.\ M.\ Fretwell, S.\ B.\ Dugdale, M.\ A.\ Alam, M.\ Biasini, L.\ Hoffmann and
A.\ A.\ Manuel,
\href{http://dx.doi.org/10.1209/0295-5075/32/9/012}
{Europhys.\ Lett.\ {\bf 32}, 771 (1995)}.

\bibitem{gull1984}
S.\ F.\ Gull and J.\ Skilling,
\href{http://dx.doi.org/10.1049/ip-f-1:19840099}
{IEE Proc.-F {\bf 131}, 646 (1984)}.

\bibitem{nordgren1989}
J.\ Nordgren, G.\ Bray, S.\ Cramm, R.\ Nyholm, J.-E.\ Rubensson and N.\
Wassdahl,
\href{http://dx.doi.org/10.1063/1.1140929}
{Rev.\ Sci.\ Instrum.\ {\bf 60}, 1690 (1989)}.

\bibitem{secondorder}
The second order emission features satisfy the second order of diffraction of
the spectrometer diffraction grating, and therefore appear at half the energy of
the first order emission features (Zn $L_3$ = 1011.7; Zn $L_2$ = 1034.7).

\bibitem{balakrishnan2004}
G.\ Balakrishnan, O.\ A.\ Petrenko, M.\ R.\ Lees and D.\ McK.\ Paul,
\href{http://dx.doi.org/10.1088/0953-8984/16/29/L02}
{J.\ Phys.: Condens.\ Matter {\bf 16}, L347 (2004)}.

\bibitem{chiuzbaian2008}
S.\ G.\ Chiuzb\u{a}ian, T.\ Schmitt, M.\ Matsubara, A.\ Kotani, G.\
Ghiringhelli, C.\ Dallera, A.\ Tagliaferri, L.\ Braicovich, V.\ Scagnoli,
N.\ B.\ Brookes, U.\ Staub and L.\ Patthey,
\href{http://dx.doi.org/10.1103/PhysRevB.78.245102}
{Phys.\ Rev.\ B {\bf 78}, 245102 (2008)}.

\bibitem{miyasaka2003}
S.\ Miyasaka, Y.\ Okimoto, M.\ Iwama and Y.\ Tokura,
\href{http://dx.doi.org/10.1103/PhysRevB.68.100406}
{Phys.\ Rev.\ B {\bf 68}, 100406(R) (2003)}.

\bibitem{benckiser2008}
E.\ Benckiser, R.\ R\"{u}ckamp, T.\ M\"{o}ller, T.\ Taetz, A.\ M\"{o}ller, A.\
A.\ Nugroho, T.\ T.\ M.\ Palstra, G.\ S.\ Uhrig and M.\ Gr\"{u}ninger,
\href{http://dx.doi.org/10.1088/1367-2630/10/5/053027}
{New J.\ Phys.\ {\bf 10}, 053027 (2008)}.

\bibitem{skilling1991}
J.\ Skilling and S.\ F.\ Gull, 
\href{http://www.jstor.org/stable/4355715}
{Inst.\ Math.\ S.\ {\bf 20}, 341 (1991)}.

\bibitem{piper2011b}
L.\ F.\ J.\ Piper, J.\ Laverock, A.\ R.\ H.\ Preston, S.\ W.\ Cho, B.\ Chen, A.\
DeMasi, K.\ E.\ Smith, L.\ Saraf, T.\ Kaspar, J.-H.\ Guo and A.\ Rusydi,
unpublished (2011).

\end{thebibliography}
\end{document}